\documentstyle[twocolumn,pra,aps,epsfig]{revtex}
\def\be{\begin{equation}}
\def\ee{\end{equation}}
\def\bea{\begin{eqnarray}}
\def\eea{\end{eqnarray}}
\begin{document}
\draft
%


\title{On the observation of decoherence with a movable mirror}
\author{R.Folman$^1$, J.Schmiedmayer$^1$ and H.Ritsch$^2$}
\address{$^1$Institut f\"ur Experimental Physik, Universit\"at Innsbruck,
Technikerstra\ss e 25, A--6020 Innsbruck, Austria\\
$^2$Institut f\"ur Theoretische Physik, Universit\"at Innsbruck,
Technikerstra\ss e 25, A--6020 Innsbruck, Austria}
\date{\today}
\maketitle
%
%
%
\begin{abstract}
Following almost a century of debate on possible `independent of
measurement' elements of reality, or `induced' elements of reality -
originally invoked as an ad-hoc collapse postulate, we propose a novel line
of interference experiments which may be able to examine the regime of
induced elements of reality. At the basis of the proposed experiment, lies the
hypothesis that all models of 'induced' elements of reality
should exhibit symmetry breaking within quantum evolution. 
The described {\em symmetry experiment} is thus aimed at being able 
to detect and resolve symmetry breaking signatures.\\
Pacs numbers: 03.65.Bz, 42.50.Vk, 42.50.Dv.
\end{abstract}
\section{Introduction}

The loss of the ability to consistently use the word particle
(referring to a classical point of mass), is of course
one of the well known implications of quantum mechanics, and stands at the
base of what has been named the `measurement problem'. Instead, we make use
of a mathematical entity called the wave function, which is allowed
superpositions which cannot describe our classical notion of reality. This
is perhaps most readily exhibited in the double slit experiment. Indeed it
was Feynman who described the double slit experiment as ''...it contains
the only mystery''\cite{Fey}. It is also a matter of general knowledge that many of
the important contributors to the theory were not satisfied with this state
of affairs. They felt that some level of independent reality does in fact
exist, and connects to quantum expectations through some set of local or
non-local hidden variables. Just to mention a few, de-Broglie for example,
tried to formulate alternatives such as the `guiding wave' or the `double
solution' models, which were, by his own admittance, unsuccessful.
Nevertheless, until his last days, he continued to believe that a theory
maintaining some sort of particle independent reality should be found \cite{deB}.
Bohm, went a step further by publishing a consistent formalism which enables
the existence of a particle, while reproducing standard quantum expectations
\cite{Bohm}. Indeed some, such as Bell, have taken the view that Bohm's success
presents a superior interpretation, while others thought differently \cite{Shimon}.
Einstein with the EPR paradox, Schr\"odinger with the cat enigma, and other
important contributors, were also uncomfortable. We refer the interested
reader to some of the many available textbooks on the interpretation of the
past and possibilities of the future regarding quantum theory \cite{Peres}.

Simply stated, the measurement problem may be described as follows: If there
are two possible pointer positions (in the measuring apparatus), the
superposition principle maintains that any superposition of those two
pointer positions must also be a possible state. However, such superposition
states of macroscopic pointers have never been observed \cite{Wigner}. 
In the language
of the above single particle double slit experiment: The superposition
principle does not allow us to use the word particle if we are to describe
the evolution of the quantum system. However, the outcome of the experiment
necessitates the use of the word particle in contradiction to the
superposition principle. Even if one accepts Bohr's escape route, which
divided the world into quantum and classical, one is left with a fuzzy,
impossible to define, border between the two.

A second class of models attempting to resolve the measurement problem
invokes `induced reality' rather than `independent reality'. Namely,
classical reality as an outcome of processes, which depend on parameters
such as time and mass (or number of particles). One example of such a model
would be the spontaneous localization through the GRW (Ghirardi, Rimini and
Weber) mechanism and its successors \cite{Sq}. Another example is what Folman and
Vager described as the `non-passive Bohmian particle' \cite{Fol}, which Bohm
described as a particle having influence on the wave function ''...so
that there will be a two way relationship between wave and particle'' \cite{Undiv}.
In this scenario, the wave function, which determines the evolution of the
system, is gradually distorted by the particle and its location -- away from
the form of superposition. There are numerous other hypothesized models such
as Gravity based induced decoherence (see Penrose \cite{Pen}), but perhaps the
most well known model of this class of `induced reality' models, is that of
a ''collapse'' due to the coupling to the environment. This is usually
referred to as {\em Decoherence}, which states in essence that the reduced
density matrix Von Neumann called for (having no off diagonal elements) may
be arrived at naturally through the entanglement of the system and the
detector to the environment (see Zurek \cite{Zurek}). In addition to the usual
parameters of time, mass and spatial separation (which are needed if we are
to explain our observations), {\em Decoherence} correlates the loss of 
coherence to the coupling onto the environment.

The question we would like to address in this note, is how may we try
and experimentally investigate the second class of models (where by
decoherence, we will be referring to the important x space), so that their
general validity would be asserted, and furthermore, how may we possibly
differentiate between them.

\section{The experimental problem}

There are two major experimental problems concerning the observation of
localization as a function of time, mass and the variation of the coupling to
the environment.\\

{\bf Problem I:}{\em It is hard to observe a quantum system without coupling 
to it and initiating unwanted decoherence as part of the measuring process. 
In such a
measurement, decoherence and the collapse postulate cannot be
differentiated, since the more we couple the environment (e.g. through our
pointer) to the observed system, the more we know about that system, and the
more dephasing we expect from the collapse postulate \cite{Sub}.
 Here, we mention
the collapse postulate in the sense of our consciousness gaining knowledge
about the system.}\\
{\bf Resolution I:}{\em Use non-demolishing measurements in which the 
system's unitary
evolution in the base of your choice (in our case it will always be x), is
not affected by the measurement.}\\

Here, it is interesting to note, that following a suggestion for a
gravitationally induced collapse model \cite{Pen}, Schmiedmayer, Zeilinger and
colleagues have also investigated the idea of monitoring the behavior of a
coherent system in order to observe decoherence \cite{Sch}. They made the point
that any `welcher weg' information would have to be erased or not invoked in
the first place, for such an experiment to be performed. As will be shown in
the following, this is exactly the idea behind our proposed experiment and
why we consider it to be non-demolishing.\\

{\bf Problem II:}{\em Traditionally, observed coherent systems in a state of 
superposition are particles, atoms or molecules. 
These are either too light to observe localization in the time
frame of the experiment, or, their mass is fixed, making it hard to
determine the proportionality of localization to mass. Furthermore,
particles in well defined states of spatial superposition,
are usually in motion, which makes the control of their
environment a hard task.}\\
{\bf Resolution II:}{\em Keep particles only as probe while turning the 
set-up into the observed system, 
which is in a well defined spatial superposition, 
with a variable mass and environment.}\\ 

Finally we note that the affect of the environment, as well as other
parameters, on localization has long been the subject of experimental
interest, but as far as we know, with no conclusive results. For example,
one such on going experimental effort concerns the handedness of chiral
molecules \cite{Zeev}. Another experiment investigated decoherence of an
`atom-cavity field' entangled state, where the macroscopic element was that
of the phase difference between the cavity fields \cite{Harosch}. Recently, several
schemes have suggested ways to directly investigate macroscopic objects
\cite{mirr1}. In this context, micro movable mirrors (which are also discussed in
this paper) have also been discussed extensively \cite{mirr2}. This, however, as far as we know, only in the context of cavities. In the following we present
what is to the best of our knowledge a novel type of experimental procedure
in the context of localization, which may shed new light on the processes
initiating it. It relates to the issue of symmetry in quantum phenomena,
which is an underlying feature of the theory. Namely that the
difference between classical and quantum states is that the phase between
possible positions is lost, and hence symmetry in space may exist between
probabilities but not amplitudes.

\section{The experimental assumptions}

In ref. \cite{Fol}, Folman and Vager proposed to incorporate the two experimental
resolutions described above by utilizing a {\em symmetry experiment} with a
movable mirror, to observe localization and decoherence. Their point was
that localization could be observed, not only by the breaking of energy
conservation (producing photon emission) -- as suggested by Pearle and
Squires \cite{Sq}, but also by the breaking of symmetry. However, they mainly
dealt with issues pertaining to unfavorable empty wave models and gave
little consideration to the experimental feasibility. In this note, we
expand the idea of the {\em symmetry experiment} to include all models of the
second type. We also include different versions of
the experiment, which may be more realizable and conclusive. Finally, we
also present initial calculations to examine the experimental
feasibility.

Before describing the experiment, we lay down the foundation by emphasizing
several assumptions that cannot, to the best of our understanding, be
avoided. For convenience, we first deal with independent collapse mechanisms
such as the GRW mechanism, and then move on to the more subtle question of
non-independent collapse mechanisms.\\

{\bf Working assumption I}: {\em The invoking of localization, via models of
the second type, destroys amplitude (wave function) symmetry, even when we
have not gained any knowledge of which of the possible x states has been
occupied. Namely, symmetric or anti-symmetric states become asymmetric.
Consequently, eigen-states of Parity are lost}.
For example, it is well known that for localized chiral molecules having well defined handedness, Parity is not a good quantum number. In general, 
if this working assumption were not valid, then it would follow that a
classical reality or at the very least the change of the wave function,
independent of our consciousness, has not been invoked by this second class
of models, although this was their main goal. Indeed, symmetry of the wave
function must be lost as the loss of the relative phase is an essential part 
of all localization schemes.\\

{\bf Working assumption II}: {\em Measuring whether Parity is a good quantum
number of the system (or even collapsing the system into an eigen-state of
Parity) does not localize the system (in the sense of the collapse
postulate), as it gives us no knowledge what so ever concerning its
location. More so, even in the context of models where localization is
independent of our knowledge, a measurement of Parity does not decohere the
system (in x space).}
This working assumption is self evident as the spatial spread of the wave
function remains unchanged by the Parity operator. 
Using once again the example of chiral molecules, measuring Parity does 
not induce handedness.
In another example, as shown by
Scully et al. \cite{Scully}, sending excited atoms through optical cavities 
which
serve as 'welcher weg' detectors, does not destroy the interference pattern if
the only knowledge gained is that a photon has been released (i.e. both
cavities were exposed to the same photon detector), and no knowledge is
available regarding at which cavity the photon has been released. As photon
emission can only be done in conjunction with a symmetric atom wave
function, the `quantum eraser' experiment shows us that measuring the Parity
of the atomic wave function, does not destroy its coherency.\\

{\bf Working assumption III}: {\em There are no other possible causes for
symmetry breaking aside from the hypothesized localization. Namely, if the
set-up is symmetric and the Hamiltonian conserves Parity (we neglect the
weak force), any sign of symmetry breaking must be due to localization.}
Again, if this was not so, we would have observed symmetry breaking long
ago, due to a breaking term in the interaction.\\

Following working assumptions I, II \& III, we set out to search for
symmetry breaking effects.

\section{The experiment}

\begin{figure}[tbp]
\begin{center}
\epsfig{file=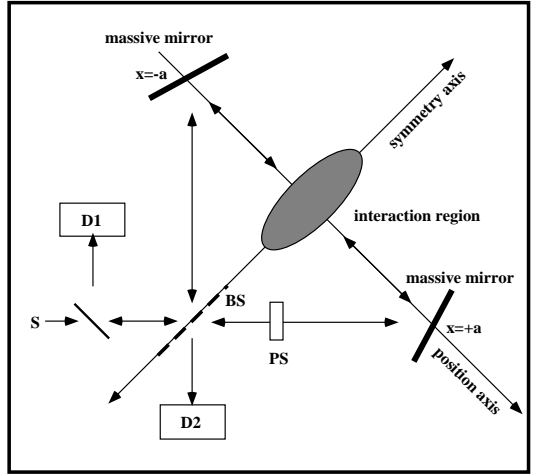,width=7cm}
\vspace*{0.5truecm}
\caption{The closed loop interferometer. As the interaction region C
is transparent, both paths are identical
as they are one of the same. S, D1, D2, BS, PS are the particle source, 
the two detectors, the beam splitter and the phase shifter, respectively.
The eigen-states of the measuring system are Parity eigen-states.
While transversing the apparatus, the initial symmetric wave function is 
exposed to the influence of the relevant parameters such as time, 
spatial separation, and environment, or any other symmetric interaction
which we place at the interaction region C. There is no reason for the
outgoing wave function to loose its initial symmetry and hence we expect
no 'click' in detector D2, unless loss of coherence due to localization
has occured.
\label{closed}} 
\end{center}
\end{figure}

The first stage of the experiment includes the preparation of a symmetric
initial particle wave function $\Phi_i$. This may be done by the apparatus 
presented in figure ~\ref{closed}. D1 and D2 are two particle detectors (if needed, with the ability to
measure the particle's energy). $S$ is the particle source. The phase shifter
($PS$) cancels the phase difference introduced by the beam splitter ($BS$), 
between the phase of the transmitted wave and that of the reflected. 
We of course assume a perfect $50\%/50\%$ $BS$ and a $PS$ which is invariant to
changes (as is the $BS$) in the wave number. The same apparatus also serves
as the measurement apparatus with Parity eigen-states.

The interaction region may hold several types of experiments.

a. {\bf The closed loop interferometer:}

This interferometer has the distinct advantage of ensuring symmetry, as its
two optical paths are one and hence identical.

In the simplest case, where we set out to examine induced collapse
independent of the environment, the interaction region may stay empty.

\noindent

The
two important parameters are time and mass (or number of elementary
particles). Mass could be controlled by the size of the particles we send
into the interferometer, and time by their velocity and the size of the
interferometer.

Dependence of localization on the environment, may be examined by
introducing a symmetric interaction which would keep the Hamiltonian
invariant to space inversion along the position axis (e.g. magnetic or
electric field or modulating crystal).

In all these cases, a 'click' in D2 would mean that the single particle
initially arriving from $S$ with a symmetric $\Phi_i$ has now transformed to 
a final $\Phi_f$ which is anti-symmetric or asymmetric. 
As there are no reasons for 
$\Phi_f$ to be anti-symmetric (see assumption III) we conclude that $\Phi_f$ 
is asymmetric. In the
single particle case some of the asymmetric photons would end up in detector
D2, telling us that the particle was localized. This could be verified by
repeating the experiment with a multi-photon pulse and observing hits on
both detectors. As the collapse postulate cannot be responsible for the
observed symmetry breaking (see assumption II), we conclude that we have
observed induced localization (see assumption I).

b. {\bf The open loop interferometer:}

\begin{figure}[tbp]
\begin{center}
\epsfig{file=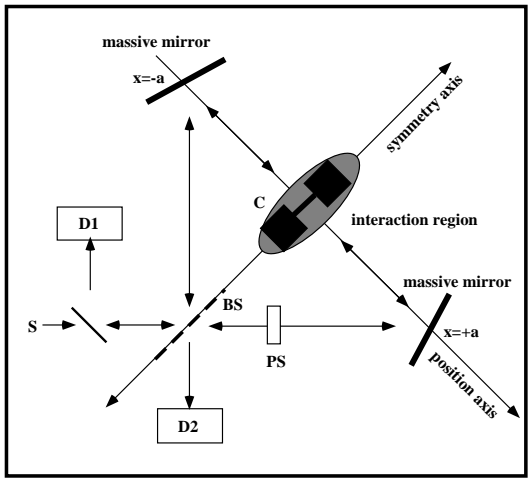,width=7cm}
\vspace*{0.5truecm}
\caption{The open loop interferometer. Here, in addition to what has been 
described in figure 1, we position a two sided foil mirror, held in an 
harmonic potential, into the interaction region C. Hence,
we entangle the transversing 
particle to the state of the set-up, which is also in a well defined quantum
state with a well defined spatial uncertainty. Thus, we are able to examine 
the quantum evolution of a macroscopic system. Using Parity eigen-states to
measure the state of the outgoing particle, we learn about the decoherence of
the set-up, without affecting it. 
\label{open}}
\end{center}
\end{figure}

If we are not able to achieve coherent symmetric states with very massive
particles (needed if we are to observe induced localization on the time
scale in which the particles transverse the apparatus), or if we are unable
to satisfactorily control the particles' environment while they are in
motion, we would have to resort to a more complex experimental scheme which
we describe here as the open loop interferometer. Different from the
previous interferometer, here, the interaction region is blocked by a second
symmetric quantum system. In this scheme, we loose the simplicity of having
only massive reflections but we gain the possibility of observing a quantum
object in a localized potential, for long periods of time and with the
ability to control its mass and environment. Consider, for example, the
set-up of figure~\ref{open}.

\noindent

Here, a mirror has been placed in the interaction region. In this example,
the mirror in the interaction region ($C$) is actually a two sided reflecting
foil which is in an harmonic oscillator potential. Neglecting inner degrees
of freedom, such as those corresponding to the Debye-Waller factors (since
it is known that these factors also exist in massive mirrors but still they
reflect coherently), we take account only of the center-of-mass of the foil,
and note that it is in a Parity eigen-state (say, the even ground state 
$\Psi_i$).
We further note that $\Psi_i$ is symmetric with respect to the same 
symmetry axis as $\Phi_i$. 
Namely, they are both symmetric with respect to the axis that lies in
the plane of the beam splitter, which is also the plane of the average
position of the foil.
%

As the total initial wave function $\Omega_i~=~\Phi_i~\otimes~\Psi_i$ is 
symmetric and as the
Hamiltonian is Parity conserving and invariant under the combined two
reflections of the x',x'' coordinates of the two wave functions, the final
total wave function $\Omega_f$
must also be symmetric and hence it may be defined as
\bea
\Omega_f~=~\sum{\Phi_s~\otimes~\Psi_s~+~\Phi_{as}~\otimes~\Psi_{as}} 
\eea
where `s' and `as' stand for symmetric and
anti-symmetric, and the summation is over all possible foil states 
(The latter is a general consequence of Parity conservation. For a
formal derivation with the specific coherent reflection Hamiltonian, 
see appendices). We note that these arguments are
independent of the mass of the foil $C$.

Changing the observation time, or the environment or the mass of the foil,
would allow us to investigate the process of induced localization.
Registering a 'click' in detector D2 with an energy that is not in
accordance with the energy gap between the foil's initial symmetric state
and one of the odd states, will indicate the occurrence of symmetry breaking
and of induced localization. Again, using a multi-photon pulse may enable a
complementary check to the single photon probe.

Let us add here that although we leave the important question of preparation
for future publications, 
we would like to note that preparing the foil in a pure 
quantum state, is for some cases not a necessary pre-condition. Any well
defined probability distribution of state occupation, will sufice. For
example, reference~\cite{mirr1} advocates the appropriateness of the thermal
state of a moving mirror, as a well defined initial quantum state.

\begin{figure}[tbp]
\begin{center}
\epsfig{file=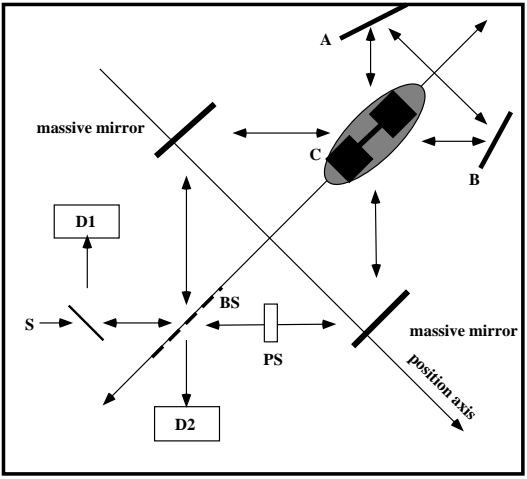,width=7cm}
\vspace*{0.5truecm}
\caption{The semi-closed loop interferometer. Here, we add two massive 
mirrors to the set-up of figure 2, so as to close the interferometer loop.
This, makes the set-up insensitive to localization signals while still 
sensitive to quantum excitations of the mirror foil, hence enabling a
further verification of our understanding of the results.
\label{semi}}
\end{center}
\end{figure}

Finally we note, that changing the set-up slightly, one may prohibit
localization signals from detector D2 (i.e. only anti-symmetric excitation
events would `click' at D2), and hence establish yet another complementary
check on the source of the deviations. As an example of such an apparatus,
we present the set-up of figure~\ref{semi}. Here, massive mirrors $A$ and $B$ 
maintain the loop actually closed.

We end this section by noting that the above presented experimental
procedure should be able to investigate any model in which decoherence is
also a function of mass and time. As these parameters are essential in all
models of the second type, we conclude that the {\em symmetry experiment} 
may be utilized to investigate the full range of models of this class.

\noindent

\section{Some numbers}

Let us make some initial calculations for the open loop interferometer, to
show what kind of experimental procedure would be needed to realize this
experiment.

We start by the demand that the wavelength $\lambda$ of the probing 
particle would
be much smaller than the localization position $X_{loc}$
which is of the order of the
width $W$ of the ground state at interaction region $C$ 
(If this were not so, the localization signal would be suppressed as 
$\frac{I_2}{I_1}=\tan^2(4\pi\frac{W}{\lambda})$ 
where $I_1$ and $I_2$ are the photon
intensities at detectors D1 and D2. A derivation is given in appendix A). For
now, let us then assume an harmonic potential so that 
$W=\sqrt{\hbar(n+\frac{1}{2})/(\frac{1}{2}m\omega)}>>\lambda$
(which only differs by a factor 
$\sqrt{2}$ from the frequently used $W=\sqrt{<x^2>}$).

\begin{figure}[tbp]
\begin{center}
\epsfig{file=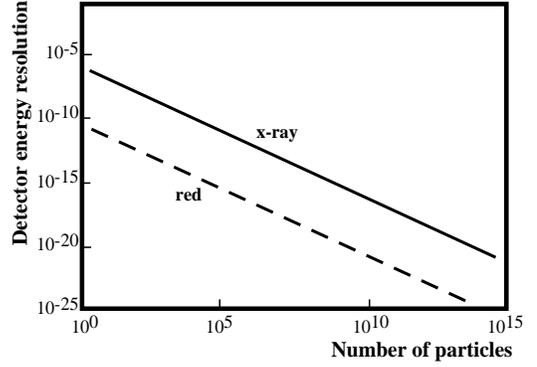,width=7cm}
\vspace*{0.5truecm}
\caption{Needed energy resolution $\frac{\hbar\omega}{E_\gamma}$
as a function of the number of foil 
nucleons, for the two examined extremes of the feasibly
used light spectrum.
\label{res}}
\end{center}
\end{figure}

We would first like to see if we are able to isolate through energy
measurements of the outgoing particles, those `clicks' at D2 which do not
originate from the standard odd excitations of the foil mirror. 
If we take for example the photon as a probing particle, and assume that we
are able to use light from x-ray to red at a wavelength of
$0.1-1000nm$, and system $C$ is say in the ground $n=0$ state, we find that for
x-ray (red): 
\bea
\sqrt{\frac{\hbar}{m\omega}}>>0.1nm (1000nm)~or~ 
m\omega<<10^{-14} (10^{-22})
\eea 
The needed energy resolution $\frac{\hbar\omega}{E_\gamma}$
in order to be able to take
account only of non excitation events, as a function of the number of
particles in the foil, is presented in figure ~\ref{res}.

\noindent 

Demanding a detector energy resolution of say $10^{-13}$, 
the harmonic frequency $\omega$ should 
be less than $10^5$, which would mean for the preferred x-ray, an
oscillating mass of $10^{-19}kg$ or $10^8$ particles.
This would mean a $10^{-7}$ collapse probability per second (If one is to
take for example a GRW factor of $10^{-15}$ collapse probability per
particle per second \cite{Bell}). To observe ten events one would need three 
parallel set-ups continuously performing a one second experiment for the 
length of a year. Indeed, it seems that with present available energy 
detection resolution, it would be very hard to perform this experiment, 
while maintaining sensitivity to the full range of possible second class 
models and parameter values~\cite{Limit}.

Following the above conclusion, let us now examine whether perhaps energy
measurements would not be needed since excitations will be suppressed by the
maximally allowed energy transfer. We note that the maximum energy transfer
by the photon cannot exceed that allowed by momentum conservation, namely:
$\frac{2(\hbar k)^2}{m}\approx\frac{8\times 10^{-47}}{m}$
which would mean in the case of a $10^8$ nucleon oscillator, an approximate 
transfer of $8\times 10^5Hz$. This should be compared with the 
$10^4Hz$ maximally allowed energy spacing, under condition (2). 
This ratio between the
maximum transferred energy by momentum conservation and the maximally
allowed harmonic energy spacing is constant for all masses and wavelengths, 
and has the value of $\frac{2(\hbar k)^2}{m}/\frac{\hbar^2}{m\lambda^2}=8\pi^2$. Namely, momentum conservation does not prohibit the harmonic oscillator
system from being excited to at least the first odd level. 

As we have taken $|k_i|$ and $|k_f|$ to be identical, 
the latter calculation is of course only valid in the
limit of massive objects which due to their mass do not receive significant
recoil energy.  
For exact numbers, we would need to make the quantum
calculation which is simple enough. The probability for the foil not to be
excited is simply the well known Debye-Waller factor $P_{0\rightarrow 0}$
which is for the case of the x-ray, and again under condition (2), smaller than
\bea 
\exp(-\hbar^2(2k)^2/2m\hbar\omega)=\exp(-8\hbar\pi^2/\lambda^2~10^{-14})\approx10^{-35}
\eea
(The full calculation may be found in appendix
B). We see that the energy transfer can actually be equal to many times the
harmonic energy spacing. Hence, we come to the conclusion that working
within the `high resolution' $W>>\lambda$
regime and with non-excitation events, is not
feasible (this is true for both x-ray and red light).
 
Finally, one may then ask: Can we avoid the need for energy measurement 
altogether by simply preparing an experimental procedure in which 
$P_{0\rightarrow n(odd)}$ (i.e. the
expected excitation signal at D2) is well known? Namely, by knowing what the
expected `noise' in D2 is (i.e. the anti-symmetric photons coming from
excitation events), one can differentiate between the signal at D2 and the
`noise'. However, for the experiment to be feasible, one must make sure that
the noise does not overwhelm the signal. Hence, we need to calculate the
values of the ratio 
\bea
R=\frac{1-P_{0\rightarrow n=0,2,4,..}}{I_2/(I_2+I_1)}
\eea
where, 
\bea
\frac{I_2}{I_2+I_1}=\int_{-\infty}^{+\infty} f(X_{loc}) \sin^2(4\pi\frac{X_{loc}}{\lambda}) dX_{loc}
\eea
and where, 
\bea
f(X_{loc})=\Psi_{0}^\dagger(X_{loc})\Psi_0(X_{loc})
\eea

One more possibility to bypass the need for identifying the photons via an
energy measurement, is to use multi-photon pulses, as described before.

In the following section we investigate these two propositions.

\section{The proposed experimental procedure}

From the previous section it is understood that for lack of very small line
width sources and very high resolution energy detectors, it seems we are
left with two viable experimental procedures, which do not make use of
energy measurements:

a. Single photon experiment, where the search will be for deviations from
the calculated ratio $I_1/I_2$ 
between symmetric and anti-symmetric photon final
states. Here, many repetitions of the same single photon experiment will
enable us to directly measure the ratio between symmetric and anti-symmetric
final states, which in turn is a consequence of the excitation
probabilities. This experiment will enable us to measure the excitation
probabilities in a quantum macroscopic system and to be sensitive to
deviations which are due to localization.

b. Multi-photon (pulse) experiment, where $I_1/I_2$ are 
different for a symmetric
outgoing pulse, an anti-symmetric outgoing pulse and an a-symmetric outgoing
pulse. Here, a localization event may be detected by the abnormal intensity
of a single pulse split at the beam splitter.

Experiment a: To see if this experimental procedure allows 
for a reasonable signal over noise ratio, 
and for what values of the parameters it does so, we calculate $R$
as a function of $m,~\lambda$ and $\omega$. 
As mentioned before, we ignore the internal degrees
of freedom (based on the coherent reflection of visible light mirrors and
x-ray gratings) and treat the center-of-mass motion of the mirror. Namely,
as in the M\"ossbauer effect, there are no excitations of internal degrees of
freedom. Whether or not this is a good approximation, depends on the
specifics of the incoming wave and the mirror, including its thickness and
material (see in the following, discussion regarding the extinction
coefficient in section `The mirror'). With the above assumption, and in the
Lamb-Dicke limit (see full calculation in appendix C), we find $R$ to be 
\bea
\frac{\eta^2+(higher~orders)}{I_2/(I_2+I_1)}
\eea
where $\eta$ is the Lamb-Dicke parameter $\sqrt{\frac{E_r}{\hbar\omega}}=\frac{4\pi}{\sqrt{2}} W/\lambda$ ($E_r$ is the recoil energy), 
and where we have neglected factors which appear in appendix C.

\begin{figure}[tbp]
\begin{center}
\epsfig{file=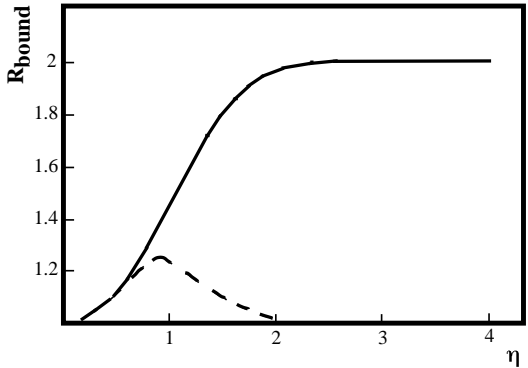,width=7cm}
\vspace*{0.5truecm}
\caption{$R_{bound}$ as a function of $\eta$, which in turn is a function of
the foil mass and frequency, 
and the wavelength of the light, with the relation 
$\eta=\frac{4\pi}{\sqrt{2}} W/\lambda=\frac{4\pi}{\lambda}\sqrt{\hbar/2m\omega}$.
The dashed line qualitatively presents the general expected form of $R$ 
(see text).
Note that the Lamb-Dicke limit is only valid for $\eta<<1$. 
\label{rbound}}
\end{center}
\end{figure}

Of course, whether or not a system is in the Lamb-Dicke limit depends on
the chosen experimental parameters. Indeed, the accurate calculation of $R$,
which is essential for any experimental realization, will depend heavily on
the set-up. Hence, keeping to a general frame, we plot in figure~\ref{rbound}
a simple example of the behavior of $R$: 
the upper bound of $R$, namely, 
\bea
R_{bound}=\frac{1-P_{0\rightarrow 0}}{I_2/(I_2+I_1)}
\eea
(this simple enough calculation is also presented in appendix C).
The rather striking `inversion' in $R_{bound}$ seen in the figure,
is dependent on the mass of the
foil, its frequency and the frequency of the impinging light, only
via the parameter $\eta$! For x-ray and 
$10^{15}$ particles, the frequency of the foil is 
$f=\frac{10^{-1}}{\eta^2}Hz$. 
This means that in order to observe the
inversion in $R$, one would have to work in the regime of very low 
frequencies. However, it is expected that the full 'inversion' presented
in figure~\ref{rbound}, would not be experimentally observable 
in any case; as can
be readily seen from the Debye-Waller expression in eq. (3),
which simply equals $\exp(-\eta^2)$: 
$P_{0\rightarrow 0}<10\%(>90\%)$ for $\eta>1.5(<0.3)$. 
One may then roughly assume that for the first case 
$1-P_{0\rightarrow 0}\rightarrow 2(1-P_{0\rightarrow n=0,2,4,..})$ since 
$P_{0\rightarrow 0}\rightarrow 0$ 
and $P_{0\rightarrow 0,2,4,..}\rightarrow \frac{1}{2}$.
Similarly for the second case,
$1-P_{0\rightarrow 0}\rightarrow 1-P_{0\rightarrow n=0,2,4,..}$ since
$P_{0\rightarrow 0}\rightarrow 1$ 
and $P_{0\rightarrow 0,2,4,..}\rightarrow 1$.
Hence, we find respectively
that for $\eta>1.5$: $R\rightarrow \frac{1}{2}R_{bound}$, 
and for $\eta<0.3$: $R\rightarrow R_{bound}$. Consequently, this
means that instead of observing the full 'inversion' one should be able to 
observe a peak in the value of $R$ for $\eta$ in the region of $0.9$.
This is qualitatively described in figure~\ref{rbound}.

\noindent 

As $R\leq~R_{bound}$, 
the experimental sensitivity for the various values of the experimental
parameters, will be better than that indicated by the bound. Observing the
above form of the dependence of $R$ as a function of $m$, 
may serve as
yet another verification that the deviations that are observed at detector
D2, indeed originate from localizations. For example, for a constant $10Hz$
foil frequency, the range of $0.1<\eta<2$
could be scanned by simply changing the mass of
the foil in the range $10^{12}<no.~of~particles<10^{15}$.
 
As for red light, one can easily see that for $10^{15}$ particles 
$f=\frac{10^{-9}}{\eta^2}Hz$, which means that although one
may perform the experiment with red light (reasonable anti-symmetric `noise'
to signal ratio), one is not able to explore the `inversion' regime, as the
needed foil frequency would be far below $1Hz$ (which would probably be
mechanically hard to achieve. See following section: ''The mirror''), or
alternatively, the needed foil mass would be extremely small, and in order
to observe decoherence, one would have to maintain a stable and isolated
experiment, for long periods of time. Nevertheless, if one has enough
statistics, one may try and observe the logarithmic behavior of $R$ for smaller
values of the $\eta$ parameter. In any case,
it should be once again noted, that as the observation of the `inversion' or
logarithmic behavior are mere verifications, the experiment itself could be
performed with a large range of light and mirror frequencies.

Finally, we add that although in general, localization times in the different
models depend on the number of particles or the mass, some models also take 
into account the spatial seperation. For our bound mirror foil, the larger
the mass, the smaller the ground state size and hence the position 
uncertainty, which constitutes the seperation between the different possible
positions. Thus increasing the mass will on the one hand shorten the 
decoherence time but on the other hand prolong it. It is therefore clear that
the exact parameters needed in order to achieve decoherence on the 
experimental time scale, are model dependent. However, it is also clear that
any model attempting to explain the 'localization of the pointer' must also
predict the localization of our mirror. It remains to adjust the 
experimental parameters so that the predicted decoherence times are within
the experimental time scales.

We now turn to the second possible experiment.

Experiment b: The immediate and clear advantage of a 
multi-photon pulse experiment, is of
course the fact that a single experiment (i.e. one pulse) can detect a
localization. The clear signal would be an $I_2/I_1$
ratio which is classically related to the $X_{loc}$ of the localization. 
Integrating over all observed
localization signals, should of course be in agreement with the density
function of the ground state. The important issue at hand is: what would be
the signal coming from non-localization events (coherent excitation and
non-excitation events).

Let us assume the pulse as being a simple sum of $N$ independent photons. Let
us also assume that we are working in the limit where $N\times P_{int}<<1$, 
where $P_{int}$
is the probability of an inelastic (i.e. excitation) photon-foil
interaction. Namely, if a photon inelastically interacts with the foil, the
chance of another such interaction in that pulse, is negligible. As we have
seen for the Lamb-Dicke limit, this is indeed the case. In this limit we can
expect that a maximum of one photon would 'click' at D2 for every pulse, while
for a localization event, we would expect on average many more. We will
leave further elaboration regarding this option for a specific note, and
simply state that as long as the above assumptions are correct, and as long
as the pulse time imitates the single photon experiment in that it is short
relative to the period of the oscillator, we expect the multi-photon
experiment to have interesting features worth examining.

We now turn to discuss some aspects of the mirror design.

\section{The mirror}

In the previous paragraphs, we calculated the excitation probability without
taking into account the specific features of the mirror. Obviously, a full
account of the mirror features has to be made. This is not just a matter of
the mirror's material. The mirror's thickness may also dominate the ability
of the mirror to reflect or to have a wanted ground frequency (e.g. of the
order of the example $10Hz$). In the following we present some preliminary
classical considerations which will have to be followed when constructing
the mirror.

For example, if one takes an atom to have a volume of $10$ cubic Angstrom,
then a $1 mm^2$ mirror, would have a thickness of only $10-1000$ 
atom layers for $10^{13}$ to $10^{15}$ atoms. 
The question then arises if such a small thickness can have
non-negligible reflectance. Taking perpendicular impinging beams (i.e.
parallel and perpendicular polarisations give rise to the same border
reflection), and for simplicity ignoring interference between reflections
coming from the two mirror bounderies, one should expect reflection from
each of them, with a strength of $|(N_1-N_2)/(N_1+N_2)|^2$ 
where $N_1$ and $N_2$ are the complex indices of
refraction of the mirror and its surroundings ($N=n+i\kappa$, 
where $n$ is the index of refraction and $\kappa$
is usually referred to as the extinction coefficient. If $n/\kappa>>1$,
then $\kappa$ may be neglected in the above calculation). 
From the latter it is clear
that in order to reflect, the mirror material must be immersed in an 
environment with an index of refraction different than its own, or
alternatively, when this is hard to achieve such as for x-ray, the mirror
should be constructed to Bragg reflect the light. 
$\kappa$ also affects the internal scattering and heating of the mirror. 
A small $\kappa$ ensures that the internal absorption $1-\exp(-2k\kappa x)$
will remain small (where $k$ is the wave vector of the incoming wave
and $x$ is simply the propagation distance of the wave within the material).
Experimental values for the extinction coefficient in the x-ray region are
usually of the order of $10^{-6}$ to $10^{-7}$ ~\cite{Palik}. 
Making use of these numbers and a mirror
thickness of $100A$, one finds an absorption of less than $1\%$. Of course, a
more accurate account should also take into consideration factors like the
dependence of the extinction coefficient on temperature, errors rising from
roughness and contamination of the mirror surface, etc. \cite{Palik}. 
For red light,
having an extinction coefficient of about $10$ (for example, metals in the red
region), one finds that the absorption for a $100A$ mirror would be above 
$50\%$ and must therefore be seriously considered.

One should also consider the mechanical properties of such a mirror. Namely,
can the mirror be fabricated to have the example frequency of $10Hz$. It is
well known that rectangular or circular plates with clamped edges have a
fundamental $(0,0)$ mode frequency of order $C_Lh/L^2$
(times $1.654$ or $0.4694$ for rectangular and circular, respectively), 
where $L$ is the dimension of an
oscillating plate of thickness $h$ and $C_L=\sqrt{E/\rho (1-\nu^2)}$
is the velocity of sound in the plate \cite{sound}. 
$E$ is Young`s Modulus. Taking for example metals, $E$ is in the order of
$10-20\times 10^3N/m^2$, $\rho$ is the density, 
which for metals is in the order of $10-20\times 10^3kg/m^3$, 
and $\nu$ is Poisson`s
ratio, which is $0.3$ for most materials. Taking the thickness to be $100A$ and
the mirror to be of dimension $1mm$, one finds that the frequency will be of
the order of $10Hz$ as required.

Finally , we discuss environmentally induced decoherence.

\section{Observing environmental Decoherence}

Before the end of this paper, we would like to revisit the issue of the work
assumptions. The working assumptions, which have been presented in the
beginning of the paper, form the underlying logic behind the hypothesis that
the {\em symmetry experiment} should exhibit sensitivity to all models of the
second class. However, the different models of this class have slightly
different features and hence require slightly different working assumptions to
ensure the sensitivity of the experiment to them. In the beginning of the
paper, we started for simplicity with the GRW model. As an example of a
different model, of a non-independent collapse nature, we now briefly
discuss the environmentally induced collapse model. Here we will simply
repeat the three working assumptions for the case of environment related
collapses, and note that a fourth assumption should be added:

{\bf Working assumption I:} {\em The invoking of coherence-loss, via models of the second type, destroys amplitude (wave function) symmetry, even when we have
not gained any knowledge of which of the possible x states has been
occupied. Namely, symmetric or anti-symmetric states become asymmetric.
Consequently, eigen-states of Parity are lost.}
Here we have replaced the original mention of localization with the more
general and perhaps appropriate word `coherence-loss'.
If this working assumption were not valid, then it would follow that a
classical reality or at the very least the loss of phase relations,
independent of our consciousness, has not been invoked by the {\em Decoherence}
model, although this was its main goal \cite{Dec}.

{\bf Working assumption II:} {\em No change.}

{\bf Working assumption III:} {\em No change.}

Finally, if our experiment is to be able to observe {\em Decoherence}, 
we need a fourth assumption:

{\bf Working assumption IV:} {\em For the case of Decoherence, we assume that 
for two counter-propagating particles (in the case of the closed-loop 
interferometer) or for two counter-propagating foil states (in the 
case of the open-loop interferometer), 
the states of the environment are orthogonal. 
Hence, in the framework of Decoherence, the main condition for
coherence-loss to occur, is fulfilled by the experiment we have described.}
As one can see in reference~\cite{Zurek}, the states of the environment
corresponding to a certain two-state superposition, must be orthogonal
if decoherence is to occur. 
If this condition cannot be realized by any kind of 
'aconscious' interaction between our system and the environment,
just as it would for two pointer positions,
then again, as in assumption I, we find that the model has not achieved what
it has set out to do, namely, arrive at a loss of coherence independent of
our consciousness.

\section{Outlook}
 
In a sequential paper, we will specifically treat the predictions of the
different models of the second type in the context of the {\em symmetry
experiment}. There, we will also address the question of system preparation
(e.g. initial mirror states~\cite{mirr1} and mirror cooling~\cite{Mancini}).

One should of course also thoroughly examine the mirror model and other 
realistic mechanisms which are able to produce an asymmetric signal, e.g.
collisions of background gases with the mirror. The latter can, for example,
be isolated through different scaling laws with respect to the mirror surface
area.

Finally, one should note that other, perhaps advantageous, possibilities for 
the interaction region $C$, may include large atoms or
molecules or perhaps even condensate gases, in a trap with very long coherence
times.

\section{Summary and conclusion}

We have discussed the class of induced localization models, among them
{\em Decoherence}. We have shown that if these models comply with several
assumptions, essential to their philosophy, then, a symmetry based
experiment should be able to investigate the hypothesized loss of coherency.

\acknowledgements

One of us (R.F.) is grateful for enlightening discussions with Professors
Anton Zeilinger, Zeev Vager, Yakir Aharonov and Jeffrey Bub. We are also
thankful for helpful comments made by Markus Gangl.
 
\appendix

\section{Localization signal}

Let us assume the foil $C$ is localized at distance $|X_{loc}|$ from the $x=0$ 
symmetry axis.

Let us now assume a normal plane wave $\exp(ikx)$ where $k$ is the absolute 
value of the wave number. 
Dividing away the normalization and other identical factors, and
remembering that the phase shifter cancels the phase difference introduced 
by the beam splitter (here for example we take $\pi/2$), one finds for
$\frac{I_1}{I_2}$:
\begin{center}
$\frac{|\exp(i[k(x+2X_{loc})])+\exp(i[k(x-2X_{loc})])|^2}{|\exp(i[k(x+2X_{loc})+\pi/2])+\exp(i[k(x-2X_{loc})+3\pi/2])|^2}=$\\
$\frac{|[\exp(i2kX_{loc})+\exp(-i2kX_{loc})]\exp(i\pi)|^2}{|\exp(i2kX_{loc})\exp(i\pi/2)+\exp(-i2kX_{loc})\exp(i3\pi/2)|^2}=$\\
$\frac{\cos^2(2kX_{loc})}{\sin^2(2kX_{loc})}$
\end{center}
where, for simplicity, we have neglected taking account of the expected 
non-negligible transmittance of the mirror, due to its small thickness.
 
One should also consider the fact that in a bound state it could be that
localizations would cause excitation to higher quantum levels with specific
Parity, rather than ending up as the localized asymmetric states we are
considering here~\cite{Sq}. Hence, in a real experiment, this rate should
be calculated and subtracted from the signal.

\section{Debye-Waller factor}

Let us calculate the Debye-Waller factor for the foil:

First, we expand the ground state in the momentum basis 
$|i\rangle=\sum_{k'}|k'\rangle\langle k'|i\rangle$ and note that 
$\exp(-ik^{\Delta}x)$
operating on a plane wave state, changes the wave number by an amount 
$k^{\Delta}$ which
is simply the difference between the incoming photon wave number $k_1$
and that of the outgoing photon $k_2$ i.e. 
$k^{\Delta}=k1+k2$ namely,
$\exp(ik^{\Delta}x)|k'\rangle=|k'-k^{\Delta}\rangle$. 
Summing over all $k'$ one gets the familiar 
$|f\rangle=\exp(-ik^{\Delta}x)|i\rangle$. Now,
\bea 
P_{0\rightarrow n}=|\int \Psi_{0}^{\dagger}(x)(K_{+}+K_{-})\Psi_n(x)|^2
\eea
where $K_{+}=\frac{1}{2}(\exp(-ik^{\Delta}x)+\exp(ik^{\Delta}x))$ and
$K_{-}=\frac{1}{2}(\exp(-ik^{\Delta}x)-\exp(ik^{\Delta}x))$
are simply the symmetric and anti-symmetric kick operators. 
The factor $\frac{1}{2}$ 
in the $K$ operators, which comes
from the normalization of the photon wave function, ensures that although
different from the standard M\"ossbauer calculation i.e. here we have a kick
from both sides, the result stays the same (see appendix C).

Let us now calculate $P_{0\rightarrow 0}$. 
We note $Q_n(x)= \Psi_{n}^{\dagger}(x)\Psi_n(x)$  
as the density operator and find:
\bea
P_{0\rightarrow 0}=|\int \Psi_{0}^{\dagger}(x)K_{+}\Psi_0(x)|^2\approx\nonumber
\eea
\bea
|\int Q_0\frac{1}{2}(2-(k^{\Delta})^2x^2)|^2=|\int Q_0-\frac{(k^{\Delta})^2}{2}\int Q_0x^2|^2=\nonumber
\eea
\bea
|1-\frac{(k^{\Delta})^2}{2}\langle x^2 \rangle _0|^2\approx 1-(k^{\Delta})^2\langle x^2\rangle _0\approx\nonumber
\eea
\bea
\exp(-(k^{\Delta})^2\langle x^2\rangle _0)=\exp(-\hbar^2(k^{\Delta})^2/2m\hbar\omega)
\eea
As for this case $k_1=k_2$ since there is no energy transfer, the final result
is
\bea
\exp(-2\hbar^2k_{1}^{2}/m\hbar\omega)
\eea
The above probability for the foil not to be excited is simply the well
known Debye-Waller factor for the case of reflection. 

\section{Excitation probabilities}

In the previous appendix, we presented the classical quantum calculation for 
the Debye-Waller factor. In the following, we present the same calculation but
in the language of annihilation and creation operators, in a way which can be
easily expanded to calculate excitation probabilities to all levels.
Furthermore, in this appendix we rigorously describe how the system Hamiltonian
allows for excitations to both symmetric and anti-symmetric states.

Let us consider $H$ the total coherent scattering Hamiltonian of our system 
$H_{light}+H_{c.m.-mirror}+H_{polarization}+H_{interaction}$
to be: 
\bea
\hbar\nu(\hat a_{+}^{\dagger}\hat a_{+}+\hat a_{-}^{\dagger}\hat a_{-}+1)+\nonumber
\eea
\bea
\hbar\omega(\hat b^{\dagger}\hat b+\frac{1}{2})+\hbar\mu(\hat c^{\dagger}\hat c+\frac{1}{2})-\alpha E^2
\eea
where, $\nu$, $\omega$ and $\mu$ 
are the frequencies of the light, mirror and polarization, respectively,
and $\hat a$, $\hat b$ and $\hat c$, 
are the usual creation and annihilation operators. 
$\hat a_{+}$ and $\hat a_{-}$ denote
photons going right and left along the x-axis of the experimental set-up
described earlier.
$\alpha$ is the polarizability and $E$ is the electric field of the
incoming light, which is simply: 
\bea
{\bf E} = {\bf \epsilon} \big\{ \hat a_{+}e^{(ikx)}+ \hat a_{-}e^{(-ikx)}- \hat a_{+}e^{(-ikx)}- \hat a_{-}e^{(ikx)} \big\}\tilde{e}
\eea 
where $\tilde{e}=i\sqrt{\frac{\hbar\nu}{2\epsilon_0V}}$ 
and ${\bf \epsilon}$ is the polarization vector (${\bf P}=\alpha{\bf E}$ is
usually denoted as the polarization of the medium. Here, for simplicity, we
neglect the variation of ${\bf P}$
as a function of $x$. For wavelengths short compared
to the thickness of the mirror, this will of course have to be taken into
account. We also note that expressing ${\bf P}$
in terms of $(\hat c+\hat c^{\dagger})$ as is usually done, and
using the adiabatic approximation $\frac{d\hat c}{dt}=i[H,\hat c]=0$
to calculate $\hat c$, gives the same result).
Hence we find for $H_{int}$, 
\bea
-\alpha\big\{ (2\hat a_{+}\hat a_{-}-\hat a_{+}\hat a_{+}^{\dagger}-\hat a_{-}\hat a_{-}^{\dagger}-\hat a_{+}^{\dagger}\hat a_{+}+2\hat a_{+}^{\dagger}\hat a_{-}^{\dagger}-\hat a_{-}^{\dagger}\hat a_{-})+\nonumber
\eea
\bea
(\hat a_{+}-\hat a_{-}^{\dagger})^2e^{(i2kx)}+(\hat a_{-}-\hat a_{+}^{\dagger})^2e^{(-i2kx)}\big\}\tilde{e}^2
\eea 
Noting that the first term is responsible for off-energy-shell
(virtual) photons and phase shifts, we write:
\bea
V_{eff}=\alpha\tilde{e}^2\big\{ (\hat a_{+}\hat a_{-}^{\dagger}+\hat a_{+}^{\dagger}\hat a_{-})2\cos(2kx)+\nonumber
\eea
\bea
i(\hat a_{+}\hat a_{-}^{\dagger}-\hat a_{+}^{\dagger}\hat a_{-})2\sin(2kx) \big\}
\eea
We see here, how $V_{eff}$
has the ability to excite the foil into a symmetric state
while leaving the photon wave function symmetric, or alternatively, to
excite the foil into an anti-symmetric state while changing the photon state
from symmetric to anti-symmetric. Expressing the latter formally, we note:
\bea
|\Psi(0)\rangle=\frac{1}{\sqrt{2}}|0\rangle_m (|1\rangle|0\rangle+|0\rangle|1\rangle)_p
\eea
where $m$ means `mirror' and $p$ 'photon'.
As $\frac{i}{\hbar}\frac{d}{dt}|\Psi(t)\rangle=H_0|\Psi(t)\rangle+V_{eff}|\Psi(t)\rangle$,  
the changes in the wave function will be proportional to:  
\bea
2\cos(k^{\Delta}x)|0\rangle_m \frac{1}{\sqrt{2}}(|1\rangle|0\rangle+|0\rangle|1\rangle)_p+\nonumber
\eea
\bea
2i\sin(k^{\Delta}x)|0\rangle_m \frac{1}{\sqrt{2}}(|1\rangle|0\rangle-|0\rangle|1\rangle)_p
\eea
where $k^{\Delta}$ has
been defined in the previous appendix. The relative excitation probabilities
will thus be: 
\bea
P_{0\rightarrow n(even)}=\frac{\alpha^2\tilde{e}^4}{\hbar^2}|\langle n|\cos(k^{\Delta}x)|0\rangle|^2
\eea
and
\bea
P_{0\rightarrow n(odd)}=\frac{\alpha^2\tilde{e}^4}{\hbar^2}|\langle n|\sin(k^{\Delta}x)|0\rangle|^2
\eea 

Let us calculate $P$:

In the Lamb-Dicke limit \cite{Lamb}, $\eta=\sqrt{\frac{E_r}{\hbar\omega}}=k^{\Delta}\sqrt{\frac{\hbar}{2m\omega}}=\frac{4\pi}{\sqrt{2}} W/\lambda<<1$, 
where $\eta$ is the Lamb-Dicke parameter, $W$ the size
of the harmonic potential ground state, $\lambda$ 
the wavelength of the impinging light, $m$
and $\omega$ the mass and frequency of the oscillator, 
and $E_r$ the recoil energy.
As can be readily seen, in this limit the ground state size (or recoil
energy) is much smaller than the wavelength of the incoming beam (or
oscillator energy spacing). Hence, an expansion of the excitation matrix
element in powers of $\eta$, is allowed. 
Making use of our typical numbers (i.e. $m=10^{15}$ particles and $\omega=2\pi\times10Hz$),
one finds that our system is within this limit for the full range of x-ray
to red light. This limit is of course very different from our initial
demand of $\sqrt{\frac{\hbar}{m\omega}}>>\lambda$, 
which results for the same mass and frequency range, in
extremely low values for the harmonic oscillator frequency. 
It is also very different than the parameter regime which one would need
in order to observe the described form of $R$. We now turn to
calculate the probability for excitation to the even and odd states. Using
the results of the previous appendix and the above definition of $\eta$, 
and using the normal convention for the position operator $\hat x=(\frac{\hbar}{2m\omega})^{\frac{1}{2}}(\hat a+\hat a^{\dagger})$, 
we find:
\bea
P_{0\rightarrow n(even)}=\sum_{n} |\langle n|\cos[\eta(\hat a+\hat a^{\dagger})]|0\rangle|^2\approx\nonumber
\eea
\bea
\sum_{n} |\langle n|1-\frac{1}{2}\eta^2(\hat a+\hat a^{\dagger})^2|0\rangle|^2=1-\eta^2
\eea
\bea
P_{0\rightarrow n(odd)}=\sum_{n} |\langle n|\sin[\eta(\hat a+\hat a^{\dagger})]|0\rangle|^2\approx\nonumber
\eea
\bea
\sum_{n} |\langle n|\eta(\hat a+\hat a^{\dagger})|0\langle|^2=\eta^2
\eea
where we expanded up to second order in $\eta$, 
and where all probabilities should
be multiplied by $\frac{\alpha^2\tilde{e}^4}{\hbar^2}$
and by the total photon scattering probability $\Omega$.
A very small $\Omega$ (e.g. due to small thickness), 
will cause the overall intensity in D2 to be much smaller than that calculated 
above, as $P$ are calculated only for
the portion of the photons which are scattered.

2. In order to calculate $R_{bound}$, one simply needs to calculate 
$I_2/(I_2+I_1)$, as we have already
calculated $P_{0\rightarrow 0}$ in the previous appendix.

As
\bea
\int_{0}^{\infty}\exp(-a^2x^2)\cos(bx)=\int_{0}^{\infty}\exp(-a^2x^2)(1-2\sin^2(\frac{b}{2}x))\nonumber
\eea
and as
\bea
\int_{0}^{\infty}\exp(-a^2x^2)=\frac{\sqrt{\pi}}{2a}\nonumber
\eea
and
\bea
\int_{0}^{\infty}\exp(-a^2x^2)\cos(bx)=\frac{\sqrt{\pi}\exp(-b^2/4a^2)}{2a}\nonumber
\eea
one finds:
\bea
\int_{-\infty}^{+\infty}\exp(-a^2x^2)\sin^2(\frac{b}{2}x)=\frac{\sqrt{\pi}}{2a}(1-\exp(-b^2/4a^2))\nonumber
\eea

Defining, $a=\sqrt{\frac{m\omega}{\hbar}}$ and $b=\frac{8\pi}{\lambda}$, 
we find:
\bea
R_{bound}=\frac{1-P_{0\rightarrow 0}}{I_2/(I_2+I_1)}=\nonumber
\eea
\bea
2\frac{1-\exp(-8\pi^2/\lambda^2a^2)}{1-\exp(-16\pi^2/\lambda^2a^2)}
\eea


\begin{thebibliography}{9}
\bibitem{Limit}It is unclear what are the exact dynamics with which the whole
collective 'pointer' follows the localization of one atom, and how
the localization time would be influenced by the fact that for some 
experimental parameters, 
the uncertainty in the collective position (ground state size of the mirror) 
is smaller than $a$ which is the GRW spatial spread of the collapsed wave
function. It could very well be that for some experimental parameters, and
for some specified dynamics, even smaller decoherence times of the collective
are to be expected. 
On the other hand, we note for example, that some models, 
such as the Continuous
Spontanious Localization model~\cite{Sq}, predict much shorter collapse times.
For such models, it could very well be that the symmetry experiment may be
performed using availbale energy detection resolution.
\bibitem{Dec}See for example: ''Conscious observers have lost their monopoly...The
environment can also monitor a system$\ldots $ such monitoring causes
decoherence, which allows the familiar approximation known as classical
objective reality...'' in ref. [11].
\end{thebibliography}
\end{document}